\documentclass[aps,prl,twocolumn,showpacs]{revtex4}

\usepackage{epsf}

\begin{document}

\title{Atomic States Entanglement in Carbon Nanotubes}

\author{I.V. Bondarev}
\altaffiliation{Also at: Institute for Nuclear Problems,
Belarusian State University, Bobruiskaya Str.11, 220050 Minsk,
Belarus}
\author{B. Vlahovic}
\affiliation{Physics Department, North Carolina Central
University, 1801 Fayetteville Str, Durham, NC 27707, USA}

\begin{abstract}
A scheme for entangling atoms (ions) located close to or
encapsulated inside a carbon nanotube is investigated using the
photon Green function formalism for quantizing electromagnetic
fields in the presence of quasi-one-dimensional absorbing and
dispersing media. Small-diameter metallic nanotubes are shown to
result in a high degree of the quantum bit entanglement for
sufficiently long times.
\end{abstract}

\maketitle

In spite of impressive experimental demonstrations of basic
quantum information effects in a number of different mesoscopic
systems, such as quantum dots in semiconductor microcavities, cold
ions in traps, nuclear spin systems, atoms in optical resonators,
Josephson junctions, etc., their concrete implementation is still
at the proof-of-principle stage (see, e.g., Ref.~\cite{Brandes}
for a review). The development of materials that may host quantum
coherent states with long coherence lifetimes is a critical
research problem for the nearest future~\cite{Lukin}.~There is a
need for the fabrication of quantum bits (qubits) with coherence
lifetimes at least three-four orders of magnitude longer than it
takes to perform a bit flip.~This would involve entangling
operations, followed by the nearest neighbor interaction over
short distances and quantum information transfer over longer
distances. It is thus of vital importance to pursue a variety of
different strategies and approaches towards physically
implementing novel non-trivial applications in modern
nanotechnology.

In this Letter, we study a scheme for entangling atoms (ions)
located close to or encapsulated inside a~carbon nanotube
(CN).~CNs are graphene sheets rolled-up into cylinders of
approximately one nanometer in diameter.~Extensive work carried
out worldwide in recent years has revealed intriguing physical
properties of these novel molecular scale
wires~\cite{Dai}.~Nanotubes have been shown to be useful for
miniaturized electronic, mechanical, electromechanical, chemical
and scanning probe devices and as materials for macroscopic
composites~\cite{Dai,Baughman}. Recent experiments on the
encapsulation of single atoms (ions) into single-walled
CNs~\cite{Jeong} and their intercalation into single-wall CN
bundles~\cite{Duclaux}, along with the progress in growth
techniques of centimeter-long small-diameter single-walled
CNs~\cite{Zheng}, stimulate the study of dynamic quantum coherent
processes in atomically doped CNs.

It was shown recently that the relative density of photonic states
(DOS) and the atom-vacuum-field coupling, respectively, near CNs
effectively increase due to the presence of additional surface
photonic modes coupled to CN electronic quasiparticle
excitations~\cite{Bondarev02}.~In small-diameter CNs the strong
atom-field coupling may
occur~\cite{Bondarev04,Bondarev05,Bondarev06}
--- the property that is known to facilitate the entanglement of
spatially separated qubits~\cite{Cirac,Dung}.~Qualitatively, in
terms of the cavity quantum electrodynamics (QED), the coupling
constant of an atom (modelled by a two-level system with the
transition dipole moment $d_{A}$ and frequency $\omega_{A}$) to a
vacuum field is given by $\hbar g\!=\!(2\pi
d_{A}^{2}\hbar\omega_{A}/\tilde{V})^{1/2}$ with $\tilde{V}$ being
the effective volume of the field mode the atom interacts with
(see, e.g., Ref.~\cite{Andreani}).~For the atom (ion) encapsulated
into the CN of radius $R_{cn}$, $\tilde{V}\!\sim\!\pi
R_{cn}^{2}(\lambda_{A}/2)$ that is $\sim\!10^2$~nm$^3$ for CNs
with diameters $\sim\!1$~nm in the optical range of
$\lambda_{A}\!\sim\!600$~nm.~Approximating
$d_{A}\!\sim\!er\!\sim\!e(e^{2}/\hbar\omega_{A})$~\cite{Davydov},
one obtains $\hbar g\!\sim\!0.3$~eV. On the other hand, the
"cavity" linewidth is given for $\omega_{A}$ in resonance with the
cavity mode by $\hbar\gamma_{c}\!=\!6\pi\hbar
c^{3}/\omega_{A}^{2}\xi(\omega_{A})\tilde{V}$, where $\xi$ is the
atomic spontaneous emission enhancement/dehancement (Purcell)
factor~\cite{Andreani}.~Taking into account large Purcell factors
$\sim\!10^{7}$ close to CNs~\cite{Bondarev02}, one arrives at
$\hbar\gamma_{c}\!\sim\!0.03$~eV for 1~nm-diameter CNs in the
optical spectral range.~Thus, for the atoms (ions) encapsulated
into small-diameter CNs the strong atom-field coupling condition
$g/\gamma_{c}\gg1$ is supposed to be satisfied, giving rise to the
rearrangement ("dressing") of atomic levels and formation of
atomic quasi-one-dimensional (1D) cavity
polaritons~\cite{Bondarev04,Bondarev05,Bondarev06}.~The latter
ones are similar to quasi-0D excitonic polaritons in quantum dots
in semiconductor microcavities~\cite{Reithmaier}, that are
currently being considered a possible way to produce the excitonic
qubit entanglement~\cite{Hughes}.~Here we suggest an alternative
way to generate the qubit entanglement by using quasi-1D atomic
polariton states in CNs.~We show that small-diameter metallic
nanotubes result in sizable amounts of the two-qubit atomic
entanglement.~Envisaged applications of our scheme range from
quantum information transfer over long distances (centimeter-long
distances, as a matter of fact, since centimeter-long
small-diameter single-walled CNs are now technologically
available~\cite{Zheng}) to novel sources of coherent light emitted
by dopant atoms in CNs.

We use our previously developed photon Green function formalism
for quantizing electromagnetic field close to quasi-1D absorbing
and dispersing media~\cite{Bondarev06,Bondarev05}.~Representing
such a medium, the (achiral) CN is considered to be an infinitely
long, infinitely thin, anisotropically conducting cylinder.~Its
(axial) surface conductivity is taken to be that given by the
$\pi$-electron band structure in the tight-binding approximation
with the azimuthal electron momentum quantization and axial
electron momentum relaxation taken into account.~Two identical
two-level atoms, $A$ and $B$, are supposed to be positioned at
their respective equivalent places $\mathbf{r}_{A,B}$ close to the
CN. We assign the orthonormal cylindric basis
$\{\mathbf{e}_{r},\mathbf{e}_{\varphi},\mathbf{e}_{z}\}$ in such a
way that $\mathbf{e}_{z}$~is directed along the nanotube axis and,
without loss of generality,
$\mathbf{r}_{A}\!=\!r_{A}\mathbf{e}_{r}\!=\!\{r_{A},0,0\}$,
$\mathbf{r}_{B}\!=\!\{r_{B},\varphi_{B},z_{B}\}$.~The atoms
interact with a quantum vacuum electromagnetic field via their
transition dipole moments that are assumed to be directed along
the CN axis, $\mathbf{d}_{A,B}\!=\!d_{z}\textbf{e}_{z}$.~The
contribution of the transverse dipole moment orientations is
suppressed because of the strong depolarization of the transverse
field in an isolated CN (the so-called dipole antenna
effect~\cite{Jorio}).~We also assume the atoms to be located far
enough from each other, to simplify the problem by ignoring the
interatomic Coulomb interaction.~The total secondly quantized
Hamiltonian of the system is then given
by~\cite{Bondarev06,Bondarev05}
\begin{eqnarray}
\hat{H}&=&\int_{0}^{\infty}\!\!\!\!\!d\omega\,\hbar\omega\!\int\!d\mathbf{R}
\,\hat{f}^{\dag}(\mathbf{R},\omega)\hat{f}(\mathbf{R},\omega)+\!\!\!
\sum_{i=A,B}\!\!{\hbar\tilde{\omega}_{i}\over{2}}\,\hat{\sigma}_{iz}\nonumber\\
&+&\sum_{i=A,B}\int_{0}^{\infty}\!\!\!\!\!d\omega\!\int\!d\mathbf{R}\;[\,
\mbox{g}^{(+)}(\mathbf{r}_{i},\mathbf{R},\omega)\,\hat{\sigma}^{\dag}_{i}\label{Htwolev}\\
&-&\mbox{g}^{(-)}(\mathbf{r}_{i},\mathbf{R},\omega)\,\hat{\sigma}_{i}\,]\,
\hat{f}(\mathbf{R},\omega)+\mbox{h.c.},\nonumber
\end{eqnarray}
where the three items represent the \emph{medium-assisted}
(modified by the presence of the CN) electromagnetic field, the
two-level atoms, and their interaction with the medium-assisted
field, respectively.~The operators
$\hat{f}^{\dag}(\mathbf{R},\omega)$ and
$\hat{f}(\mathbf{R},\omega)$ are the scalar bosonic field
operators defined on the CN surface assigned by the radius-vector
$\mathbf{R}=\!\{R_{cn},\phi,Z\}$ with $R_{cn}$ being the radius of
the CN.~These operators create and annihilate the single-quantum
bosonic-type electromagnetic medium excitation of the frequency
$\omega$ at the point $\mathbf{R}$ of the CN surface.~The Pauli
operators, $\hat{\sigma}_{i}\!=\!|L\rangle\langle u_{i}|$,
$\hat{\sigma}^{\dag}_{i}\!=\!|u_{i}\rangle\langle L|$ and
$\hat{\sigma}_{iz}\!=\!|u_{i}\rangle\langle
u_{i}|\!-\!|L\rangle\langle L|$ with $i\!=\!A\mbox{ or }B$,
describe the atomic subsystem and electric dipole transitions
between its two states, upper $|u_{i}\rangle$ with either of the
two atoms in its upper state and lower $|L\rangle$ with both atoms
in their lower states, separated by the transition frequency
$\omega_{A}$.~This (bare) frequency is modified by the diamagnetic
($\sim\!\!\mathbf{A}^{2}$) atom-field interaction yielding the new
\emph{renormalized} transition frequency
$\tilde{\omega}_{i}=\omega_{A}[1-2/(\hbar\omega_{A})^{2}\!
\int_{0}^{\infty}\!d\omega\!\int\!d\mathbf{R}|\mbox{g}^{\perp}
(\mathbf{r}_{i},\mathbf{R},\omega)|^{2}]$ in the second term of
the Hamiltonian.~The dipole atom-field interaction matrix elements
are given by
$\mbox{g}^{(\pm)}\!=\!\mbox{g}^{\perp}\pm(\omega/\omega_{A})\mbox{g}^{\parallel}$
where
$\mbox{g}^{\perp(\parallel)}\!=\!-i(4\omega_{A}/c^{2})\sqrt{\pi\hbar\omega\,\mbox{Re}\,
\sigma_{zz}(\omega)}\,\,d_{z}^{\;\,\perp(\parallel)}G_{zz}(\mathbf{r}_{i},\mathbf{R},\omega)$
with $^{\perp(\parallel)}G_{zz}(\mathbf{r}_{i},\mathbf{R},\omega)$
being the $zz$-component of the transverse (longitudinal) Green
tensor (with respect to the first variable) of the electromagnetic
subsystem and $\sigma_{zz}(\omega)$ representing the CN surface
axial conductivity.~The matrix elements
$\mbox{g}^{\perp(\parallel)}$ have the property of\linebreak
$\int\!\!d\mathbf{R}|\mbox{g}^{\perp(\parallel)}
(\mathbf{r}_{i},\mathbf{R},\omega)|^{2}\!\!=\!
(\hbar^2\!/2\pi)(\omega_{A}/\omega)^{2}\Gamma_{0}(\omega)
\xi^{\perp(\parallel)}(\mathbf{r}_{i},\omega)$ with
$\xi^{\perp(\parallel)}(\mathbf{r}_{i},\omega)\!=\!
\mbox{Im}^{\perp(\parallel)}G_{zz}^{\,\perp(\parallel)}
(\mathbf{r}_{i},\mathbf{r}_{i},\omega)/\mbox{Im}G_{zz}^{0}(\omega)$
being the transverse (longitudinal) local photonic DOS functions
and $\Gamma_{0}(\omega)\!=\!8\pi\omega^{2}d_{z}^{2}\,
\mbox{Im}G_{zz}^{0}(\omega)/3\hbar c^{2}$ representing the atomic
spontaneous decay rate in vacuum where
$\mbox{Im}\,G_{zz}^{0}(\omega)\!=\!\omega/6\pi c$ is the vacuum
imaginary Green tensor $zz$-component.~The
Hamiltonian~(\ref{Htwolev}) involves only two standard
approximations: the electric dipole approximation and the
two-level approximation. The rotating wave approximation commonly
used is not applied, and the diamagnetic term of the atom-field
interaction is not neglected (as opposed to, e.g.,
Refs.~\cite{Bondarev02,Bondarev04}).

For single-quantum excitations, the time-dependent wave function
of the whole system can be written as
\begin{eqnarray}
|\psi(t)\rangle&=&\sum_{i=A,B}\!\!C_{\displaystyle{u_{i}}}(t)\,
e^{-i(\tilde{\omega}_{i}-\overline{\omega})t}|u_{i}\rangle|\{0\}
\rangle\label{wfunc}\\
&+&\int_{0}^{\infty}\!\!\!\!\!\!\!d\omega\!\!\int\!\!d\mathbf{R}\,
C_{L}(\mathbf{R},\omega,t)\,e^{-i(\omega-\overline{\omega\!}\,)t}
|L\rangle|1(\mathbf{R},\omega)\rangle,\nonumber
\end{eqnarray}
where $\overline{\omega}=\!\sum_{i=A,B}\tilde{\omega}_{i}/2$,
$|\{0\}\rangle$ is the vacuum state of the field subsystem,
$|\{1(\mathbf{R},\omega)\}\rangle$ is its excited state with the
field being in the single-quantum Fock state, $C_{\displaystyle
u_{i}}$ and $C_{L}$ are the respective probability amplitudes of
the upper states and the lower state of the system. For the
following it is convenient to introduce the new variables
$C_{\pm}(t)\!=\![C_{\displaystyle u_{A}}(t)\pm C_{\displaystyle
u_{B}}(t)]/\sqrt{2}$ that are the expansion coefficients of the
wave function~(\ref{wfunc}) in terms of the maximally entangled
2-qubit atomic states
$|\pm\rangle\!=\!(|u_{A}\rangle\pm|u_{B}\rangle)/\sqrt{2}$. In
view of Eqs.~(\ref{Htwolev}) and (\ref{wfunc}), the time-dependent
Schr\"{o}dinger equation yields then
\begin{equation}
\mbox{\it\.{C}}_{\pm}(\tau)=\int_{0}^{\tau}\!\!\!d\tau^{\prime}
K_{\pm}(\tau-\tau^{\prime})\,C_{\pm}(\tau^{\prime})+f_{\pm}(\tau),\label{popampu}
\end{equation}
where
\begin{equation}
~\!\!\!\!\!K_{\pm}(\tau-\tau^{\prime})\!=\!-\!\!\int_{0}^{\infty}\!\!\!\!\!\!dx
\frac{\tilde{\Gamma}_{0}(x)}{2\pi}\xi^{\pm}(\mathbf{r}_{A},\mathbf{r}_{B},x)
e^{-i(x-\tilde{x}_{A})(\tau-\tau^{\prime})}\!,\!\!\label{Kpm}
\end{equation}\vspace{-0.5cm}
\begin{eqnarray}
\xi^{\pm}(\mathbf{r}_{A},\mathbf{r}_{B},x)\!\!&=&\!\!\frac{x^{2}_{A}}{x^{2}}\left[
\xi^{\perp}(\mathbf{r}_{A},x)\pm\xi^{\perp}(\mathbf{r}_{A},\mathbf{r}_{B},x)\right]\label{ksipm}\\
&+&\!\!\xi^{\parallel}(\mathbf{r}_{A},x)\pm\xi^{\parallel}(\mathbf{r}_{A},\mathbf{r}_{B},x),\nonumber\\
f_{\pm}(\tau)\!\!&=&\!\!-\frac{1}{\sqrt{2}}\int_{-\Delta\tau}^{0}\!\!\!\!\!d\tau^{\prime}
K_{\pm}(\tau-\tau^{\prime})\,C_{\displaystyle{u_{A}}}(\tau^{\prime}),
\label{fpm}
\end{eqnarray}
and we have, for convenience, introduced the dimensionless
variables $\tilde{\Gamma}_{0}\!=\!\hbar\Gamma_{0}/2\gamma_{0}$,
$x\!=\!\hbar\omega/2\gamma_{0}$, and $\tau\!=\!2\gamma_{0}t/\hbar$
with $\gamma_{0}\!=\!2.7$~eV being the carbon nearest neighbor
hopping integral ($\hbar/2\gamma_{0}\!=\!1.22\times\!10^{-16}$~s)
appearing in the CN surface axial conductivity $\sigma_{zz}$.~The
functions $f_{\pm}(\tau)$ are only unequal to zero when the two
atoms are initially in their ground states, with the initial
excitation residing in the nanotube.~Eq.~(\ref{fpm}) assumes that
this situation is realized by selecting the time origin to be
right after the time interval $\Delta\tau$, that is necessary for
the (excited) atom $A$ to decay completely into the nanotube
photonic modes, has elapsed.~The two-particle local photonic DOS
functions
$\xi^{\perp(\parallel)}(\mathbf{r}_{A},\mathbf{r}_{B},x)\!=\!
\mbox{Im}^{\perp(\parallel)}G_{zz}^{\,\perp(\parallel)}
(\mathbf{r}_{A},\mathbf{r}_{B},x)/\mbox{Im}G_{zz}^{0}(x)$ are the
generalizations of the DOS functions
$\xi^{\perp(\parallel)}(\mathbf{r}_{A},x)$ (see
Refs.~\cite{Bondarev06,Bondarev05}).~They are of the form of
$\xi^{\,\perp}\!=\!1+\overline{\xi}^{\,\perp}$,
$\xi^{\parallel}\!=\!\overline{\xi}^{\parallel}$ with the
medium-dependent contributions given for $r_{A,B}\!<\!R_{cn}$ by
\begin{equation}
\left\{\!\!\begin{array}{c}
\overline{\xi}^{\,\perp}(\mathbf{r}_{A},\mathbf{r}_{B},x)\\
\overline{\xi}^{\,\parallel}(\mathbf{r}_{A},\mathbf{r}_{B},x)
\end{array}\!\!\right\}={3\over{\pi}}\,\mbox{Im}\!\!\!
\sum_{p=-\infty}^{\infty}\!\!\!e^{ip\varphi_{B}}\!\!\int_{C}\!\!dy
\left\{\!\!\begin{array}{c}v^{4}\\y^{2}v^{2}\end{array}\!\!\right\}
s(R_{cn},x)\label{ksiperpar}
\end{equation}\vspace{-0.5cm}
\[
\times{K_{p}^{2}[vu(R_{cn})x]I_{p}[vu(r_{A})x]I_{p}[vu(r_{B})x]
\cos[u(z_{B})xy]\over{1+s(R_{cn},x)v^{2}I_{p}[vu(R_{cn})x]K_{p}[vu(R_{cn})x]}}\,,
\]
where $I_{p}$ and $K_{p}$ are the modified cylindric Bessel
functions, $v(y)\!=\!\sqrt{y^{2}-1}\,$,
$u(r)\!=\!2\gamma_{0}r/\hbar c$, and $s(R_{cn},x)\!=\!2i\alpha
u(R_{cn})x\overline{\sigma}_{zz}(R_{cn},x)$ with
$\overline{\sigma}_{zz}\!=\!2\pi\hbar\sigma_{zz}/e^{2}$ being the
dimensionless CN surface conductivity per unit length and
$\alpha\!=\!e^{2}/\hbar c\!=\!1/137$ representing the
fine-structure constant.~The integration contour $C$ goes along
the real axis of the complex plane and envelopes the branch points
$y\!=\!\pm1$ of the function $v(y)$ in the integrands from below
and from above, respectively.~For $r_{A,B}\!>\!R_{cn}$,
Eq.~(\ref{ksiperpar}) is modified by the replacement
$r_{A,B}\!\leftrightarrow\!R_{cn}$ in the Bessel function
arguments in the numerator of the integrand.

The entanglement of two quantum bits occurs when the 2-qubit wave
function cannot be represented as a~product of the two 1-qubit
states in any basis.~To determine this quantity in our particular
case, we follow the recipe based on the "spin flip" transformation
and valid for an arbitrary number of qubits (see
Ref.~\cite{Wooters} for details).~First, we define the reduced
density matrix
$\hat{\rho}_{AB}(\tau)\!=\!|\psi_{AB}(\tau)\rangle\langle\psi_{AB}(\tau)|\!
=\!\mbox{Tr}_{field}|\psi(\tau)\rangle\langle\psi(\tau)|$
describing the bipartite atomic subsystem in terms of the wave
function~(\ref{wfunc}) of the whole system. Next, we introduce the
"concurrence"
$\mbox{C}(\psi_{AB})\!=\!|\langle\psi_{AB}|\tilde{\psi}_{AB}\rangle|$
where $|\tilde{\psi}_{AB}\rangle\!=
\!\hat{\sigma}_{y}^{A}\hat{\sigma}_{y}^{B}|\psi_{AB}^{\ast}\rangle$
with $\hat{\sigma}_{y}^{A(B)}$ being the Pauli matrix that
represents the "spin flip" transformation in the atom $A(B)$
single-qubit space.~This, after some algebra, becomes
$\mbox{C}[\psi_{AB}(\tau)]\!=\!|C_{+}^{2}(\tau)-C_{-}^{2}(\tau)|$
with $C_{\pm}(\tau)$ given by the integral
equation~(\ref{popampu}).~Finally, the degree of the entanglement
of the 2-qubit atomic state $|\psi_{AB}\rangle$ is given by
$\mbox{E}[\psi_{AB}(\tau)]\!=\!h\{1+\sqrt{1-\mbox{C}[\psi_{AB}(\tau)]^{2}}/2\}$
where $h(y)\!=\!-y\log_{2}y-(1-y)\log_{2}(1-y)$.

The entanglement $\mbox{E}[\psi_{AB}(\tau)]$ is maximal when the
coefficients $C_{+}(\tau)$ and $C_{-}(\tau)$ are maximally
different from each other. For this to occur, the functions
$\xi^{\pm}(\mathbf{r}_{A},\mathbf{r}_{B},x)$ in
Eqs.~(\ref{popampu})--(\ref{fpm}) should be different in their
values.~These are determined by
$\overline{\xi}^{\,\perp(\parallel)}(\mathbf{r}_{A},\mathbf{r}_{B},x)$
whose frequency behavior is determined by the CN surface axial
conductivity $\sigma_{zz}$. We computed
$\overline{\xi}^{\,\perp(\parallel)}$ from Eq.(\ref{ksiperpar})
with $\sigma_{zz}$ calculated beforehand in the relaxation-time
approximation (relaxation time
$3\!\times\!10^{-12}$~s~\cite{Tans}) at temperature
300~K~\cite{Bondarev04,Bondarev05,Bondarev06}. Then, the integral
equation was solved numerically to obtain $C_{\pm}(\tau)$ and
$\mbox{E}[\psi_{AB}(\tau)]$. The vacuum spontaneous decay rate was
estimated from the expression
$\tilde{\Gamma}_{0}(x)\!\approx\!\alpha^{3}x$ valid for atomic
systems with Coulomb interaction~\cite{Davydov}. The atomic
transition frequency $x_{A}$ was assumed to be $\sim\!x_{r}$, the
local DOS resonance frequency. The simplest, most interesting case
was considered where both of the atoms are positioned in the
center of the CN.

\begin{figure}[t]
\epsfxsize=8.65cm\centering{\epsfbox{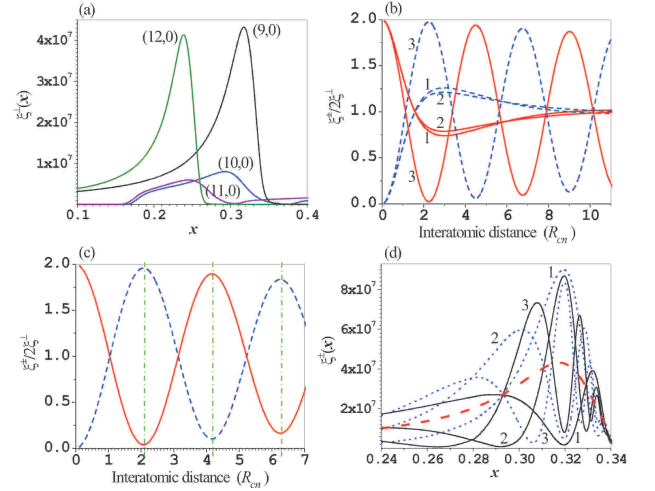}} \caption{(Color
online) \textbf{(a)}~Transverse local photonic DOS functions
$\xi^{\perp}(x)$ for the two-level atom in the centers of the four
'zigzag' nanotubes. \textbf{(b)}~Normalized two-particle local
photonic DOS functions $\xi^{+}$ (solid lines) and $\xi^{-}$
(dashed lines) taken at the peak frequencies of $\xi^{\perp}(x)$
[see (a)], as functions of the distances between the two atoms on
the axes of the (10,0) (lines~1; $x\!=\!0.29$), (11,0) (lines~2;
$x\!=\!0.25$) and (12,0) (lines~3; $x\!=\!0.24$) CNs.
\textbf{(c)}~Same as in (b) for the two atoms on the axis of the
(9,0) CN ($x\!=\!0.32$). \textbf{(d)}~Two-particle DOS functions
$\xi^{+}(x)$ (solid lines) and $\xi^{-}(x)$ (dotted lines) for the
two atoms located in the center of the (9,0) CN and separated from
each other by the distances of
$2.1R_{cn}\!\approx\!7.4\,$\AA\space (lines~1),
$4.2R_{cn}\!\approx\!14.8\,$\AA\space (lines~2) and
$6.3R_{cn}\!\approx\!22.2\,$\AA\space (lines~3) [shown by the
vertical lines in (c)]; the dashed line shows $\xi^{\perp}(x)$ for
the atom in the center of the (9,0) CN.} \label{fig1}
\end{figure}

Shown in Fig.~\ref{fig1}~(a) is a typical frequency behavior of
the one-particle transverse photonic DOS
$\xi^{\perp}(\mathbf{r}_{A}\!=\!0,x)$ in the infrared and visible
frequency range $x<0.4$ for the atom inside 'zigzag' CNs of
increasing radii. The DOS resonances are seen to be much sharper
for metallic CNs ($m\!=\!3q,~q\!=\!1,2,...$) than for
semiconducting ($m\!\ne\!3q$) in agreement with the fact that this
frequency range is dominated by the classical Drude-type
conductivity which is larger in metallic
CNs~\cite{Bondarev02}.~Figure~\ref{fig1}~(b) shows the normalized
two-particle local photonic DOS functions $\xi^{\pm}$ taken at the
resonance frequencies of the respective $\xi^{\perp}$'s
[Fig.~1~(a)], as functions of the distance between the two atoms
on the axes of the (10,0), (11,0) and (12,0) CNs.~The values of
$\xi^{+}$ and $\xi^{-}$ are seen to be substantially different
from each other before they reach their limit values of
$\xi^{+}=\xi^{-}=\xi^{\perp}+\xi^{\parallel}\approx2\xi^{\perp}$
at sufficiently large interatomic separations.~For metallic CNs
the functions $\xi^{+}$ and $\xi^{-}$ exhibit resonator-like
behavior, i.e. they vary periodically in antiphase almost without
damping with increasing interatomic separation. In
Fig.~\ref{fig1}~(c) is shown the dependence of
$\xi^{\pm}(x\!=\!0.32)$ [peak position of $\xi^{\perp}$ in
Fig.~\ref{fig1}~(a)] on the interatomic separation for the atoms
in the center of the metallic (9,0) CN. As the separation
increases, $\xi^{+}$ and $\xi^{-}$ may differ greatly one from
another in a periodic way.~The maximal difference is
$|\xi^{+}-\xi^{-}|\approx2(\xi^{\perp}+\xi^{\parallel})
\approx4\xi^{\perp}\!\sim\!10^{8}$, making the mixing coefficients
$C_{\pm}(\tau)$ different and thus resulting in a substantial
degree of the entanglement of the two spatially separated atomic
qubits.~Figure~\ref{fig1}~(d) gives an example of the $\xi^{\pm}$
frequency behavior at $x\!\sim\!x_{r}\!=\!0.32$ for the three
interatomic distances [shown by the vertical lines in
Fig.~\ref{fig1}~(c)] corresponding to "antinode" relative
positioning of the atoms in the center of the (9,0) CN.

\begin{figure}[t]
\epsfxsize=8.65cm\centering{\epsfbox{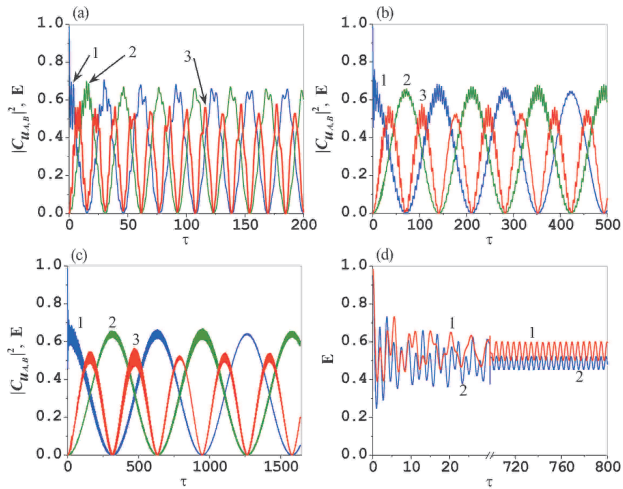}} \caption{(Color
online) \textbf{(a)},\textbf{(b)},\textbf{(c)} Upper decay of the
initially excited atom~$A$ (lines~1) and initially unexcited
atom~$B$ (lines~2), and the entanglement (lines~3), as functions
of dimensionless time for the two atoms located in the center of
the (9,0) CN and separated from each other by the distances of
$2.1R_{cn}\!\approx\!7.4\,$\AA,
$4.2R_{cn}\!\approx\!14.8\,$\AA\space and
$6.3R_{cn}\!\approx\!22.2\,$\AA, respectively [the situation shown
in terms of local photonic DOS in
Figs.~\ref{fig1}(c),(d)].~\textbf{(d)} Short-time and long-time
entanglement evolution of the initially fully entangled atoms
separated by $7.4\,$\AA; lines 1 and 2 are for $C_{+}(0)\!=\!1,
C_{-}(0)\!=\!0$ and $C_{+}(0)=0, C_{-}(0)=1$, respectively.}
\label{fig2}
\end{figure}

The ensuing spontaneous decay dynamics and atomic entanglement are
presented in Fig.~\ref{fig2} for the atoms in the center of the
(9,0) CN.~In (a),(b),(c) atom $A$ is supposed to be initially
excited while atom B is in its ground state. The entanglement is
seen to reach the amount of 0.5 and to vary with time periodically
without damping at least for the (reasonably long) times we
restricted ourselves in our computations.~As the interatomic
separation increases, so the period of the entanglement
oscillations does while no change occurs in the maximal
entanglement.~In (d) both of the atoms are initially maximally
entangled [$C_{\pm}(0)\!=\!1$ while $C_{\mp}(0)\!=\!0$] and
separated by the distance of $7.4$~\AA. The entanglement is
slightly larger in the case where $C_{+}(0)\!=\!1,C_{-}(0)\!=\!0$
and no damping occurs as before.~Note that the atoms can be
separated by longer distances with roughly the same entanglement
due to the $\xi^{\pm}$ periodicity with interatomic distance.

We have demonstrated sizable amounts of the two-qubit atomic
entanglement in small-diameter metallic CNs.~The entanglement
greatly exceeds 0.35 that is known to be the maximal value one can
achieve in the weak atom-field coupling regime~\cite{Dung}, and
persists with no damping for very long times. We attribute such a
behavior to the strong atom-field coupling and electronic
structure resulting in the resonator-like distance dependence of
the two-particle local DOS in metallic CNs. The scheme studied can
be generalized to the multi-atom entanglement via the nearest
neighbor pairwise quantum correlations, thus challenging novel
applications of atomically doped CNs in quantum information
science.

This work was supported by NSF (ECS-0631347), DoD
(W911NF-05-1-0502) and NASA (NAG3-804) grants. Discussions with
M.Gelin are gratefully acknowledged.


\begin{thebibliography}{99}
\bibitem{Brandes}T.Brandes, Phys. Rep. \textbf{408}, 315 (2005).
\bibitem{Lukin}A.S.S$\not\!\mbox{o}$rensen \emph{et al}., Phys. Rev. Lett. \textbf{92},
063601 (2004).
\bibitem{Dai}H.Dai, Surf. Sci. \textbf{500}, 218 (2002).
\bibitem{Baughman}R.H.Baughman, A.A.Zakhidov, and W.A.de Heer,
Science \textbf{297}, 787 (2002).
\bibitem{Jeong}G.-H.Jeong \emph{et al}., Phys. Rev. B \textbf{68}, 075410 (2003);
Thin Solid Films \textbf{435}, 307 (2003); M.Khazaei \emph{et
al}., J. Phys. Chem. B \textbf{108}, 15529 (2004).
\bibitem{Duclaux}L.Duclaux, Carbon \textbf{40}, 1751 (2002);
H.Shimoda \emph{et al}., Phys. Rev. Lett. \textbf{88}, 015502
(2002).
\bibitem{Zheng}S.M.Huang, B.Maynor, X.Y.Cai, and J.Liu, Advanced
Materials, \textbf{15}, 1651 (2003); L.Zheng \emph{et al}., Nature
Materials, \textbf{3}, 673 (2004).
\bibitem{Bondarev02}I.V.Bondarev \emph{et al}., Phys. Rev. Lett. \textbf{89},
115504 (2002).
\bibitem{Bondarev04}I.V.Bondarev and Ph.Lambin, Phys. Rev. B \textbf{70}, 035407 (2004);
Phys. Lett. A \textbf{328}, 235 (2004).
\bibitem{Bondarev05}I.V.Bondarev and Ph.Lambin, Phys. Rev. B \textbf{72}, 035451 (2005);
Solid State Commun. \textbf{132}, 203 (2004).
\bibitem{Bondarev06}I.V.Bondarev and Ph.Lambin, in: \emph{Trends in Nanotubes Research}
(Nova Science, New York, 2006).
\bibitem{Cirac}J.I.Cirac \emph{et al}., Phys. Rev. Lett. \textbf{78}, 3221 (1997).
\bibitem{Dung}H.T.Dung \emph{et al}., J. Opt. B \textbf{4}, S169 (2002).
\bibitem{Andreani}L.C.Andreani, G.Panzarini, and J.-M.G\'{e}rard,
Phys. Rev. B \textbf{60}, 13276 (1999).
\bibitem{Davydov}A.S.Davydov, \emph{Quantum Mechanics} (NEO, Ann
Arbor, MI, 1967).
\bibitem{Reithmaier}J.P.Reithmaier \emph{et al}., Nature \textbf{432}, 197
(2004); T.Yoshie \emph{et al}., ibid. \textbf{432}, 200 (2004);
E.Peter \emph{et al}., Phys. Rev. Lett. \textbf{95}, 067401
(2005).
\bibitem{Hughes}S.Hughes, Phys. Rev. Lett. \textbf{94}, 227402 (2005).
\bibitem{Jorio}A.Jorio \emph{et al}., Phys. Rev. B \textbf{65}, 121402R(2002).
\bibitem{Wooters}W.K.Wooters, Phys. Rev. Lett. \textbf{80}, 2245 (1998).
\bibitem{Tans}S.J.Tans \emph{et al}., Nature \textbf{386}, 474 (1997).
\end{thebibliography}
\end{document}